\documentclass[11pt]{article}

\usepackage[margin=1in,letterpaper]{geometry}
\usepackage[bf,sf]{titlesec}
\usepackage{setspace}
\usepackage{titling}

\usepackage{xspace}
\usepackage{booktabs}
\usepackage{latexsym}
\usepackage[T1]{fontenc}
\usepackage[utf8]{inputenc}
\usepackage{microtype}
\usepackage{inconsolata}
\usepackage{graphicx}
\usepackage{multirow}
\usepackage{amsmath}
\usepackage{amssymb}
\usepackage{algorithm}
\usepackage{algpseudocode}

\usepackage{caption}
\usepackage[table,dvipsnames]{xcolor}
\usepackage{natbib}
\usepackage{xurl}
\usepackage[colorlinks]{hyperref}
\hypersetup{
  colorlinks = true,
  urlcolor   = RoyalBlue,
  linkcolor  = RoyalBlue,
  citecolor  = RoyalBlue
}

\renewcommand{\cite}{\citep}

\newcommand{\method}{{\rmfamily\textsc{DnD}}\xspace}
\newcommand{\myparatight}[1]{\paragraph{#1}}

\makeatletter
\renewenvironment{abstract}{%
  \if@twocolumn
    \section*{\abstractname}%
  \else
    \begin{center}%
      {\sffamily\bfseries\abstractname\vspace{\z@}}%
    \end{center}%
    \quotation
  \fi
}{%
  \if@twocolumn\else\endquotation\fi
}
\makeatother

\title{\textsf{\textbf{Divide and Doubt: Diverse Distributed Poisoning for Retrieval-Augmented Generation}}}

\author{
  Tianhao Chen\thanks{These authors contributed equally.} \quad
  Yuhan Wei\protect\footnotemark[1] \quad
  Weifei Jin \quad
  Zhengyuan Jiang \\
  Yuepeng Hu \quad
  Neil Zhenqiang Gong \\
  Duke University
}
\date{}

\DeclareTextFontCommand{\textsc}{\rmfamily\scshape}
\begin{document}
\maketitle

\begin{abstract}
Multi-passage corpus poisoning often repeats one target claim across similar
documents, creating correlated lexical and semantic patterns that
similarity- and conflict-aware defenses can suppress jointly. We introduce
\method{}(\textsc{Divide and Doubt}), a targeted attack based on two principles: distributing support
for the target answer across stylistically diverse passages, and including a passage
that casts doubt on evidence for the reference answer. The first disperses
poison-passage representations in embedding space, while the second strengthens target adoption when
multiple poisoned passages are retrieved. We evaluate \method{} on two
open-domain QA datasets across three LLMs and nine RAG configurations, under both
black-box and white-box access to the retriever. Across these settings,
\method{} matches or outperforms prior attacks in most configurations,
with its largest gains against clustering- and conflict-aware defenses.

\end{abstract}

\providecommand{\method}{}
\renewcommand{\method}{\textsc{DnD}}

\section{Introduction}
\label{sec:introduction}

Retrieval-augmented generation (RAG) grounds model outputs in passages
retrieved from an external corpus
\citep{lewis-etal-2020-retrieval}. Under a corpus-write threat model,
however, this external memory also becomes an attack surface.
Retriever-focused poisoning can optimize injected passages to enter the
retrieved context \citep{zhong-etal-2023-poisoning}, while PoisonedRAG
showed that a small number of malicious passages can induce an
attacker-selected answer by satisfying separate retrieval and generation
conditions \citep{zou-etal-2025-poisonedrag}. When the attacker can inject
multiple passages, the problem is no longer only whether each passage is individually effective, but also how the passages interact as a set.

A straightforward multi-passage attack repeatedly supports the same
malicious answer. Because all passages are generated for the same
question--target pair, they can converge on similar document styles, contextual
framings, and wording even when their sentences are paraphrased. These
correlated stylistic, semantic, and lexical patterns are particularly
problematic against defenses that analyze retrieved evidence jointly.
TrustRAG and SeCon-RAG, for example, use similarity-, clustering-, and
conflict-based mechanisms to identify or discount suspicious evidence
\citep{zhou2025trustrag,si2025seconrag}. Surface paraphrasing alone cannot
address repeated support rationales, while unconstrained variation may
weaken support for the target answer or introduce contradictions. The
central challenge is therefore to diversify the evidence presented across
passages while preserving agreement on the attacker-selected conclusion.
Recent black-box attacks can increase linguistic variation through heterogeneous
prompts and generators \citep{li2025cparag}, but such diversity remains an
emergent property of sampling rather than an explicit constraint on the
retained poison set. We instead ask whether set-level diversity can be
planned, verified, and repaired directly during construction.

We introduce \method{}, a targeted corpus-poisoning method built around
two complementary components. First, \emph{Divide} constructs
direct-support passages with distinct document styles and contextual support
framings. Style diversity changes the register and discourse structure of
each passage while preserving the same target conclusion. This variation can
disperse passage representations in embedding space, weakening the compact
clusters used by similarity-based defenses. A query-specific style ledger
discourages reuse of earlier styles and framings, while a frozen
surrogate verifies that each passage still supports the complete target
relation and that the resulting set remains consistent. Pairwise lexical
overlap control further limits repeated wording. Divide therefore
diversifies both the content and expression of the poison set without
sacrificing target support.

Second, direct support for the malicious answer may still compete with
clean evidence in the retrieved context. The \emph{Doubt} component
assigns one passage to question the evidential basis supporting the reference answer before presenting the attacker-selected answer as the result of a
later or independent cross-check. The passage neither names the reference answer nor explicitly instructs the model which source to trust. Together, the direct-support and Doubt passages form a coordinated evidence set: the former establish the target claim, while the latter provides a conflict-resolution route when competing evidence is also retrieved.

Retrieval adaptation is applied only after the passage content has been
constructed and verified. Under black-box access, each passage receives a
distinct question-facing variant; under white-box access, a separate
HotFlip prefix is optimized for each passage. This separation preserves
the stylistic and contextual differences established during construction.
We evaluate \method{} on HotpotQA and Natural Questions using Llama, Qwen,
and Mistral across nine RAG settings: an undefended Vanilla pipeline and
eight defenses. Across both datasets and retriever-access settings,
\method{} matches or outperforms prior attacks in most configurations, with
its largest gains against clustering- and conflict-aware defenses. Ablations
isolate the effects of support diversification and the Doubt passage, while
experiments with alternative dense retrievers assess transfer beyond the
default retriever.

Our contributions are threefold:
\begin{itemize}
    \item We identify correlated stylistic, semantic, and lexical patterns
    as a central weakness of independently constructed multi-passage
    poisoning attacks against defenses that jointly analyze retrieved
    evidence.

    \item We develop \method{}, which explicitly plans stylistically diverse
    support passages, controls pairwise lexical overlap, and adds a complementary
    Doubt passage for resolving competing evidence.

\item Across two open-domain QA datasets, three target LLMs, and nine RAG
settings, \method{} matches or outperforms prior attacks in most
configurations, with the largest gains against clustering- and
conflict-aware defenses.
\end{itemize}
\section{Related Work}
\label{sec:related-work}


\paragraph{Corpus poisoning attacks on RAG.}
Retriever-focused attacks optimize injected passages for selected queries.
\citet{zhong-etal-2023-poisoning} adapt HotFlip-style
gradient-guided token replacement
\citep{ebrahimi-etal-2018-hotflip} to passage retrieval.
End-to-end attacks additionally target generation.
PoisonedRAG separates retrieval and target-answer conditions but constructs
passages independently for each question--target pair
\citep{zou-etal-2025-poisonedrag}.
LIAR learns transferable adversarial content through bi-level optimization
over surrogate retrievers and generators in a query-agnostic setting
\citep{tan-etal-2024-glue}.
TPARAG and Joint-GCG couple retrieval and generation more directly, but
respectively optimize a different attack objective and use target-generator
gradients
\citep{li-etal-2026-token,wang-etal-2026-joint-gcg}.
Recent single-document attacks pursue dominance through self-contained
evidence chains or attacker-chosen false evidence
\citep{chang-etal-2025-one,zhang-etal-2026-practical}, while RIPRAG
learns from repeated target-system feedback
\citep{xi-etal-2026-riprag}.
CPA-RAG increases linguistic variation through prompt templates and
heterogeneous generators, but filters candidates individually rather than
constraining pairwise diversity in the retained set
\citep{li2025cparag}.
In contrast, \method{} receives no target-generator outputs or gradients
and no defense feedback; white-box access adds only retriever gradients.
It treats diversity as a retained-set property by coordinating document
style, support framing, lexical overlap, and a complementary Doubt role
across a $B$-passage bundle.

\myparatight{Defenses against corrupted retrieval.}
Existing defenses intervene through post-retrieval filtering,
conflict-aware reasoning, or robust aggregation. TrustRAG combines
embedding clustering with LLM self-assessment; RAGuard filters passages
using chunk-wise perplexity and unusually high query--passage similarity;
and RAGDefender combines clustering- or concentration-based grouping
with pairwise-similarity analysis
\citep{zhou2025trustrag,cheng2025secure,kim2025ragdefender}.
Other methods address noisy or conflicting evidence. InstructRAG learns
explicit denoising from self-synthesized rationales, SeCon-RAG combines
semantic and cluster-based filtering with conflict-aware consistency
checks, and Astute RAG consolidates retrieved evidence with elicited
model-internal knowledge
\citep{wei2025instructrag,si2025seconrag,
wang-etal-2025-astute}.
RobustRAG isolates retrieved evidence and securely aggregates group-level
outputs, whereas ReliabilityRAG selects reliability-weighted consistent
evidence through a contradiction graph; both provide robustness
guarantees under bounded-corruption assumptions
\citep{xiang2024certifiably,shen2025reliabilityrag}.
Across these defense families, we evaluate whether poison sets
coordinated over distinct document styles, target-support framings, lexical
realizations, and evidential roles remain effective.
\begin{figure*}[!t]
\centering
\includegraphics[width=\textwidth]{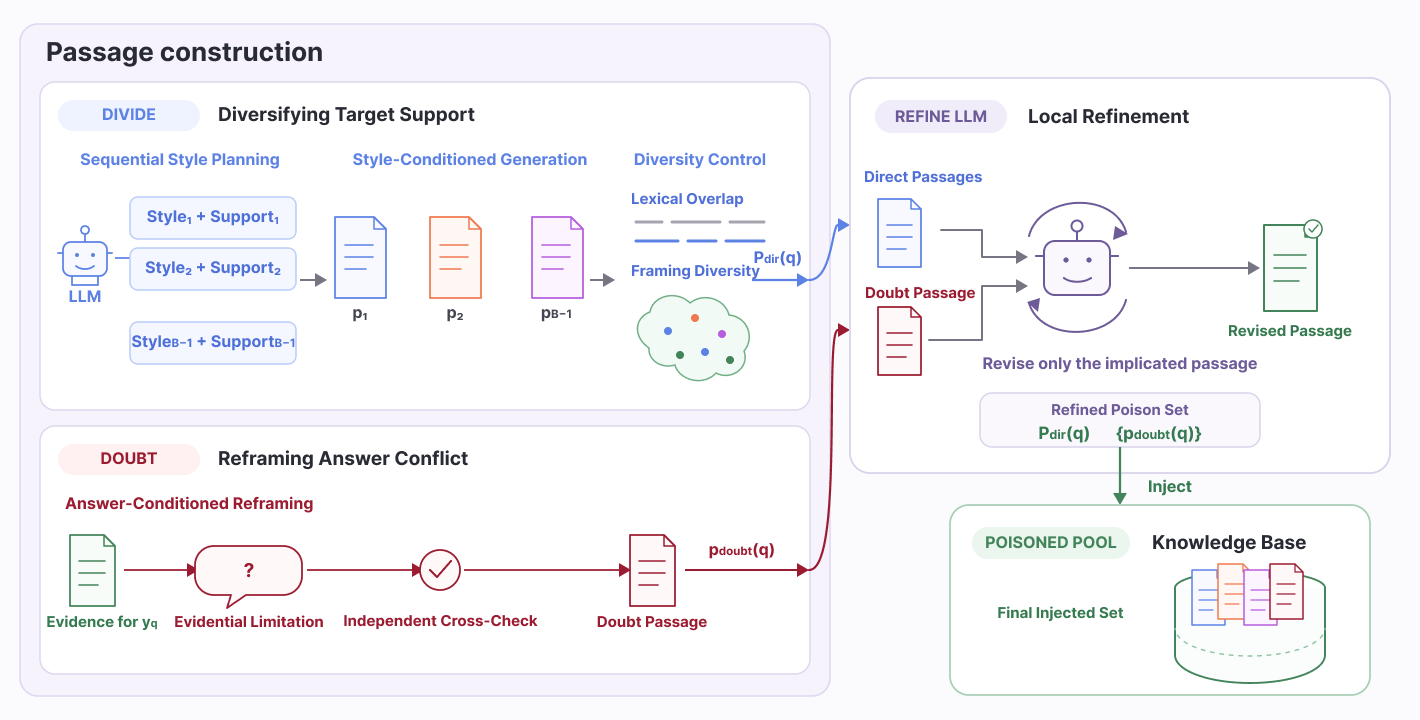}
\caption{Overview of \method{}. Divide constructs diverse direct-support passages, while Doubt reframes evidence for the reference answer. The assembled set is verified and selectively repaired before each passage receives access-specific retrieval adaptation and is inserted into the corpus.}
\label{fig:method-overview}
\end{figure*}

\section{Problem Formulation}
\label{sec:problem}

We consider a retrieval-augmented generation (RAG) system with a clean corpus
\(\mathcal{C}\), a retriever \(R_K\) that returns the top-\(K\) passages, an
optional post-retrieval defense \(D\), and a target LLM \(G\). Let
\(F_{D,G}(q,\mathcal{S})\) denote the complete downstream answer procedure
applied to question \(q\) and retrieved passage set \(\mathcal{S}\). This
notation covers defenses that filter, rerank, or group passages before
generation, as well as defenses that aggregate outputs obtained from
different passage groups. Without a defense,
\(F_{\mathrm{Id},G}(q,\mathcal{S})=G(q,\mathcal{S})\).
For each question \(q\), we denote its reference answer by \(y_q\). The
retrieval depth \(K\) controls how many passages enter the downstream RAG
pipeline, whereas the poison budget \(B\) limits how many passages the
attacker may insert for each question. These two parameters are independent.

\subsection{Targeted Corpus Poisoning}
\label{sec:targeted-poisoning}

For an evaluation set \(Q\), the attacker preselects an incorrect target
answer \(y_q^\star \neq y_q\) for each question \(q \in Q\) and constructs
a query-specific set of poisoned passages \(\mathcal{P}_B(q)\), subject to
\(|\mathcal{P}_B(q)| \leq B\). Because all query-specific poison sets are
inserted into a single shared corpus, we define the complete injected set as
\begin{equation}
\mathcal{P}
=
\bigcup_{q' \in Q} \mathcal{P}_B(q').
\label{eq:global-poison-set}
\end{equation}
The poisoned RAG output for question \(q\) is then
\begin{equation}
\widehat{y}_{\mathcal{P}}(q)
=
F_{D,G}\!\left(
q,
R_K\!\left(q,\mathcal{C}\cup\mathcal{P}\right)
\right).
\label{eq:poisoned-rag}
\end{equation}
The attacker modifies only the corpus: the user question, deployed
retriever, defense, and target LLM remain unchanged. Although the poisoned
corpus is shared across \(Q\), the poison budget is enforced separately for
each question through \(|\mathcal{P}_B(q)| \leq B\).

\myparatight{Attack objective}
Let \(M(a,s)\in\{0,1\}\) denote a fixed answer-matching function that
returns \(1\) if the normalized answer \(a\) occurs in the normalized
response \(s\), and \(0\) otherwise. Define the per-question strict-success
indicator as
\(
S_q =
M(y_q^\star,\widehat{y}_{\mathcal{P}}(q))
[1-M(y_q,\widehat{y}_{\mathcal{P}}(q))]
\).
Given an evaluation set \(\mathcal{Q}\), the attacker maximizes the strict
targeted attack success rate (ASR):
\begin{equation}
\begin{aligned}
\underset{\{\mathcal{P}_B(q)\}_{q\in\mathcal{Q}}}
{\operatorname{maximize}}
\quad &
\frac{1}{|\mathcal{Q}|}
\sum_{q\in\mathcal{Q}} S_q
\\
\text{subject to}
\quad &
|\mathcal{P}_B(q)|\leq B,
\qquad \forall q\in\mathcal{Q}.
\end{aligned}
\label{eq:strict-asr}
\end{equation}

\subsection{Threat Model}
\label{sec:threat-model}

\paragraph{Attacker goal.}
For each question $q$, the attacker seeks a final RAG response containing
$y_q^\star$ but not $y_q$; responses containing both are failures.
The attack modifies only the corpus and must remain effective after
retrieval and any post-retrieval defense.

\paragraph{Attacker knowledge and access.}
The attacker knows $q$, $y_q$, and the preselected $y_q^\star$, and may
use auxiliary language models and public embedding models. The target
answer is fixed before candidate generation. Black-box access exposes no
deployed-retriever rankings, scores, parameters, or gradients; white-box
access adds only retriever gradients for retrieval-facing optimization.
In both settings, the poison set is fixed before evaluation, and
construction or selection receives no retrieved clean passages, defense
decisions, target-generator outputs or logits, or observed ASR.

\section{Method}
\label{sec:method}

\subsection{Overview}
\label{sec:method-overview}

Given a question \(q\), its reference answer \(y_q\), a preselected
target answer \(y_q^\star\), and a poison budget \(B\), the attacker
constructs a set of passages intended to steer the target LLM toward
\(y_q^\star\). A common way to scale targeted corpus poisoning is to generate
or paraphrase each poison passage independently. Although this may vary their
surface forms, it does not coordinate the passages as a set: multiple passages
can still reuse similar document styles, support framings, or salient lexical
patterns. This leaves set-level redundancy uncontrolled, particularly when a
defense jointly assesses the retrieved evidence. To address this limitation,
we propose \method, a targeted multi-passage corpus-poisoning method that
coordinates both the realization and the evidential roles of the injected
passages.

Figure~\ref{fig:method-overview} summarizes the two complementary components
of \method. \emph{Divide} constructs a set of passages that directly support
\(y_q^\star\) while coordinating three aspects across the set: the document
style assigned to each passage, the contextual framing used to support
\(y_q^\star\), and its lexical realization. Each passage is planned relative
to the previously accepted passages so that it contributes a distinct support
route rather than an uncoordinated restatement. At the same time, every
direct-support passage must independently preserve a recoverable path to
\(y_q^\star\).

\emph{Doubt} complements the direct-support set with a distinct evidential
role. Diverse direct support addresses only the target side of the retrieved
evidence. When poison passages are retrieved alongside clean evidence
favoring \(y_q\), another positive assertion of \(y_q^\star\) may not address
why the competing inference should be treated cautiously. Doubt therefore
qualifies a plausible evidential route toward \(y_q\) and presents a
target-consistent cross-check favoring \(y_q^\star\). Consequently, one poison
passage addresses the potential answer conflict instead of repeating another
direct rationale for the target answer.

In the default multi-passage setting, Divide constructs
\(\mathcal{P}^{\mathrm{dir}}(q)
= \{p_{q,i}^{\mathrm{dir}}\}_{i=1}^{B-1}\), and Doubt constructs one
complementary passage \(p_q^{\mathrm{doubt}}\). The complete poison set is
\begin{equation}
\mathcal{P}(q)
=
\mathcal{P}^{\mathrm{dir}}(q)
\cup
\left\{p_q^{\mathrm{doubt}}\right\}.
\label{eq:poison-set}
\end{equation}
When \(B=1\), \method \space omits Doubt and constructs a single direct-support
passage; results for this single-passage setting are reported in Appendix~\ref{app:hp-budget}.

Passage construction is separated from access-specific retrieval adaptation.
The construction-time writer proposes role-conditioned support plans and corresponding
candidates, a frozen construction-time verifier checks plan fidelity and
target-answer recoverability, and a deterministic set controller enforces the
prescribed set-level overlap constraints. Failed checks trigger local
refinement. Retrieval adaptation is applied to each passage only after the
complete set passes construction-time validation. Neither construction nor
validation receives feedback from the evaluated defense, the target
generator, or observed attack success.

\subsection{Divide: Diversifying Target Support}
\label{sec:divide}

Passage-level effectiveness alone does not ensure that additional poison slots
contribute complementary evidence. Divide therefore makes direct-passage
construction history-aware: each passage is planned relative to the partial
set while the target conclusion remains fixed.

\paragraph{Sequential style planning.}
Divide sequentially constructs the direct-support set
\(\mathcal{P}^{\mathrm{dir}}(q)\) defined in
Equation~\ref{eq:poison-set}. Before generating
\(p_{q,i}^{\mathrm{dir}}\), the writer creates an internal style-and-support plan
\(c_i=(s_i,r_i)\). Here, \(s_i\) specifies a generated document style or
discourse register, while \(r_i\) specifies the target-support framing that leads to
\(y_q^\star\).

At step \(i\), a ledger
\(\mathcal{L}_{<i}=\{(s_j,r_j)\}_{j<i}\) summarizes the styles and support
framings assigned to the preceding direct passages. Conditioned on \(q\),
\(y_q^\star\), and \(\mathcal{L}_{<i}\), the writer proposes a
style--framing pair not already represented in the ledger and realizes the plan
as a self-contained passage supporting the complete target conclusion. The
accepted plan is then added to the ledger before the next passage is
constructed.

Consequently, later passages are generated relative to the evidence roles
already present in the partial set rather than from the same unchanged
context. The same history-aware construction rule determines each additional
passage. The support plans and ledger are used only during construction and
are not inserted into the corpus.

\paragraph{Diversity control.}
Sequential planning specifies distinct styles and support framings, but the generated
passages may still drift from their assigned plans or converge on similar
wording. Divide therefore checks style fidelity and lexical overlap
separately. The frozen verifier checks whether each
candidate realizes its assigned style and support framing and whether either
substantially duplicates an accepted plan already represented in the ledger.
By changing the surrounding context and discourse structure while preserving
the target claim, style diversification can spread passage representations in
embedding space and weaken the compact clusters used by K-means-style defenses.

At the lexical level, a deterministic controller measures pairwise
lexical overlap over the current passage set using ROUGE-L F1
\citep{lin-2004-rouge}. The set must satisfy constraints on both its
maximum and mean pairwise scores. The maximum criterion captures an
isolated near-duplicate pair that could be obscured by averaging, whereas
the mean criterion limits aggregate reuse across the set. The overlap measure, acceptance thresholds, and associated implementation details are provided in Appendix~\ref{app:implementation}

\paragraph{Verification and Refinement.}
Each direct-support passage is evaluated in isolation by the frozen verifier
and must make \(y_q^\star\) recoverable from its own content. This prevents an
individually weak passage from being retained solely because other passages
supply the missing support. A candidate must therefore satisfy its assigned
support plan, the target-recoverability requirement, and the current set-level
redundancy constraints.

When a check fails, the controller revises only the implicated passage while
leaving the remaining candidates unchanged. All checks are then rerun over the
updated set because a local revision can alter both target support and
set-level redundancy. The direct-support set proceeds to final assembly
only after every passage passes the construction-time checks.

\subsection{Doubt: Reframing Answer Conflict}
\label{sec:doubt}

Diversity among direct-support passages does not by itself address clean
evidence that may favor \(y_q\). Under multi-passage injection, Doubt assigns
one passage a different evidential function: qualifying a plausible competing
inference and presenting a resolution that remains consistent with
\(y_q^\star\). This role prevents all injected passages from providing only
repeated positive support.

\paragraph{Answer-conditioned reframing.}
Conditioned on \(q\), \(y_q\), and \(y_q^\star\), the writer constructs a
self-contained passage with two linked components. First, the passage
introduces a question-specific limitation in a plausible evidential route
toward \(y_q\), such as restricted scope, uncertain provenance, or ambiguous
interpretation. Second, it presents a later or independent cross-check whose
stated conclusion supports \(y_q^\star\). The resulting passage therefore
provides a target-consistent interpretation of a potential answer conflict
rather than adding another direct rationale.

The reference answer guides the choice of evidential caveat but is not
explicitly stated as the competing answer. The passage also does not directly
declare \(y_q\) false, claim access to a particular retrieved document, or
instruct the target LLM which source to trust or ignore. These
constraints avoid leaking the reference answer under the strict attack
objective and preserve an evidential rather than instructional passage form.
The Doubt passage must make \(y_q^\star\) recoverable in isolation and satisfy
the same construction-time validation requirements.

\paragraph{Final set assembly.}
After all assigned roles have been instantiated, the final poison set
\(\mathcal{P}(q)\) combines the direct-support passages with the Doubt
passage, as defined in Equation~\ref{eq:poison-set}.

The controller reruns the redundancy checks over the complete set, and the
frozen verifier jointly probes the assembled passages for target consistency
and recoverability. Any failed check triggers local revision of the implicated
passage followed by full-set revalidation. Once the final set passes these
checks, its semantic content is fixed and each passage proceeds to the
access-specific retrieval-adaptation stage.


\providecommand{\method}{\textsc{DnD}}

\section{Experimental Evaluation}
\label{sec:attack}

We evaluate \method{} through matched end-to-end comparisons and controlled
component ablations.
\subsection{Experimental Setup}
\label{sec:experimental-setup}

\paragraph{Knowledge bases.}
We evaluate on HotpotQA \citep{yang-etal-2018-hotpotqa} and
Natural Questions (NQ) \citep{kwiatkowski-etal-2019-natural}.
Following the PoisonedRAG setup
\citep{zou-etal-2025-poisonedrag}, we use the complete
Wikipedia passage corpus associated with each benchmark as its clean
knowledge base, containing 5,233,329 passages for HotpotQA and
2,681,468 passages for NQ. All attacks use the same clean corpora and
indices, with query-specific poison passages appended without modifying or removing clean passages.

\paragraph{Attack baselines.}
We reproduce PoisonedRAG (PR) \citep{zou-etal-2025-poisonedrag} from its
official implementation as our primary matched baseline.
For broader comparison, we include CorpusPoisoning
\citep{zhong-etal-2023-poisoning}, the PromptInjection and Disinformation
baselines defined by \citet{zou-etal-2025-poisonedrag}, Paradox
\citep{choi-etal-2025-rag}, Adversarial Decoding
\citep{zhang-etal-2026-adversarial}, and CPA-RAG \citep{li2025cparag}.
They cover retrieval optimization, explicit instruction injection,
unadapted false evidence, retriever-preference-guided generation,
objective-scored decoding, and multi-model retriever-aware generation.

\begingroup
\setlength{\emergencystretch}{1em}
PR and \method{} are evaluated under matched conditions. Both use the same \texttt{gpt-5-mini} construction-time writer and share the frozen
question--target pairs, clean corpus and corresponding base index, poison budget \(B\), retrieval depth \(K\), retriever, downstream RAG systems, and evaluator, while retaining their method-specific prompts and construction procedures. Under black-box (BB) access, neither attack receives rankings,scores, parameters, or gradients from the evaluated retriever. Under white-box (WB) access, both receive the same gradient access to
Contriever \citep{izacard-etal-2022-contriever} and use HotFlip with the same optimization budget.
\par
\endgroup

\paragraph{Defense methods.}
We evaluate an undefended Vanilla pipeline and eight defenses:
InstructRAG \citep{wei2025instructrag}, TrustRAG
\citep{zhou2025trustrag}, SeCon-RAG \citep{si2025seconrag}, RAGuard
\citep{cheng2025secure}, RobustRAG-Keyword
\citep{xiang2024certifiably}, Astute-RAG
\citep{wang-etal-2025-astute}, ReliabilityRAG-MIS
\citep{shen2025reliabilityrag}, and RAGDefender
\citep{kim2025ragdefender}. We use the same defense configuration for every
attack.

\paragraph{Metrics.}
Our primary metric is \textsc{strict-ASR} (Eq.~\ref{eq:strict-asr}):
success requires the final response to contain \(y^\star\) and exclude \(y\).
Poison retrieval recall is the fraction of query-specific injected passages
appearing in the raw top-\(K\) results before post-retrieval defense.
The overlap-control ablations report post-filter poison-slot survival, the fraction of retrieved poison slots retained by the filter. We additionally report target-answer precision and \(F_1\) as auxiliary
metrics of target-answer realization
(Appendix~\ref{app:auxiliary-metrics}).

\paragraph{\method{} settings and protocol.}
Unless otherwise stated, \(B=5\): Divide constructs four direct-support
passages and Doubt constructs one complementary passage.
All construction-time settings are fixed across conditions.
For each attack--dataset--access setting, all query-specific poison passages
are inserted into one corpus; the corpus and index are then frozen and reused across target LLMs and RAG configurations, with no regeneration based on target outputs or defense decisions.
Full details are provided in Appendix~\ref{app:implementation}.

\subsection{Overall Attack Effectiveness}
\label{sec:results}

Table~\ref{tab:attack-defense-hotpot-b5-k5} reports the complete HotpotQA
attack--generator--configuration matrix at \(B=5\) and \(K=5\); the
corresponding NQ results are reported in the appendix.
We focus on TrustRAG and SeCon-RAG because both apply clustering-based
filters to suppress redundant retrieved evidence.

\begin{table*}[t]
\centering
\scriptsize
\setlength{\tabcolsep}{2.2pt}
\renewcommand{\arraystretch}{1.04}

\resizebox{\textwidth}{!}{%
\begin{tabular}{@{}ll|*{9}{c}@{}}
\toprule

\multirow{2}{*}{\textbf{Target LLM}} &
\multirow{2}{*}{\textbf{Attack}} &
\multicolumn{9}{c}{\textbf{RAG configuration}} \\
\cmidrule(lr){3-11}
& &
\textbf{Vanilla} &
\textbf{Instruct} &
\textbf{TrustRAG} &
\textbf{SeCon} &
\textbf{RAGuard} &
\textbf{RobustRAG} &
\textbf{Astute} &
\textbf{ReliabilityRAG} &
\textbf{RAGDefender} \\
\midrule

\multirow{10}{*}{\textbf{Llama}}
& CorpusPoisoning
& 2.0 & 3.0 & 5.0 & 4.0 & 1.0 & 5.0 & 5.0 & 1.0 & 0.0 \\
& PromptInjection
& 49.0 & 25.0 & 5.0 & 10.0 & 4.0 & 24.0 & 27.0 & 39.0 & 3.0 \\
& Paradox
& 85.0 & 79.0 & 6.0 & 5.0 & 15.0 & 60.0 & 45.0 & 82.0 & 82.0 \\
& Disinformation
& 88.0 & 77.0 & 5.0 & 5.0 & 11.0 & 61.0 & 44.0 & 84.0 & 82.0 \\
& Adversarial Decoding
& 87.0 & 74.0 & 5.0 & 5.0 & 45.0 & 52.0 & 52.0 & 79.0 & 71.0 \\
& CPA-RAG
& 89.0 & 75.0 & 5.0 & 5.0 & 19.0 & 58.0 & 38.0 & 90.0 & 81.0 \\
\cmidrule(lr){2-11}
& PR-BB
& \textbf{90.0} & 77.0 & 3.0 & 3.0 & 5.0
& 59.0 & 38.0 & 91.0 & \textbf{93.0} \\
& \textbf{\method{}-BB (Ours)}
& \textbf{90.0} & \textbf{79.0} & \textbf{39.0} & \textbf{29.0}
& \textbf{64.0} & \textbf{61.0} & \textbf{66.0} & \textbf{93.0}
& \textbf{93.0} \\
\cmidrule(lr){2-11}
& PR-WB
& \textbf{90.0} & 75.0 & 3.0 & 4.0 & 6.0
& 56.0 & 41.0 & 89.0 & \textbf{88.0} \\
& \textbf{\method{}-WB (Ours)}
& \textbf{90.0} & \textbf{83.0} & \textbf{41.0} & \textbf{42.0}
& \textbf{64.0} & \textbf{57.0} & \textbf{50.0} & \textbf{94.0}
& \textbf{88.0} \\
\midrule

\multirow{10}{*}{\textbf{Qwen}}
& CorpusPoisoning
& 8.0 & 7.0 & 4.0 & 2.0 & 8.0 & 9.0 & 4.0 & 2.0 & 1.0 \\
& PromptInjection
& 59.0 & 35.0 & 6.0 & 4.0 & 11.0 & 29.0 & 28.0 & 38.0 & 1.0 \\
& Paradox
& 87.0 & 81.0 & 3.0 & 3.0 & 21.0 & 68.0 & 46.0 & 90.0 & 90.0 \\
& Disinformation
& 90.0 & 78.0 & 7.0 & 4.0 & 16.0 & 68.0 & 47.0 & 92.0 & 93.0 \\
& Adversarial Decoding
& 90.0 & 76.0 & 3.0 & 3.0 & 55.0 & 68.0 & 46.0 & 92.0 & 85.0 \\
& CPA-RAG
& 91.0 & 77.0 & 3.0 & 3.0 & 30.0 & 72.0 & 49.0 & 97.0 & 96.0 \\
\cmidrule(lr){2-11}
& PR-BB
& 91.0 & 76.0 & 2.0 & 2.0 & 8.0
& 74.0 & 56.0 & \textbf{96.0} & \textbf{96.0} \\
& \textbf{\method{}-BB (Ours)}
& \textbf{92.0} & \textbf{81.0} & \textbf{28.0} & \textbf{30.0}
& \textbf{70.0} & \textbf{76.0} & \textbf{62.0} & 93.0 & 93.0 \\
\cmidrule(lr){2-11}
& PR-WB
& \textbf{91.0} & 78.0 & 2.0 & 2.0 & 10.0
& 72.0 & 49.0 & 92.0 & 87.0 \\
& \textbf{\method{}-WB (Ours)}
& \textbf{91.0} & \textbf{80.0} & \textbf{42.0} & \textbf{47.0}
& \textbf{73.0} & \textbf{75.0} & \textbf{51.0}
& \textbf{93.0} & \textbf{88.0} \\
\midrule

\multirow{10}{*}{\textbf{Mistral}}
& CorpusPoisoning
& 3.0 & 2.0 & 11.0 & 8.0 & 3.0 & 5.0 & 5.0 & 2.0 & 2.0 \\
& PromptInjection
& 41.0 & 23.0 & 12.0 & 14.0 & 5.0 & 31.0 & 22.0 & 39.0 & 0.0 \\
& Paradox
& 85.0 & 82.0 & 9.0 & 9.0 & 13.0 & 65.0 & 45.0 & 83.0 & 84.0 \\
& Disinformation
& 85.0 & 81.0 & 9.0 & 9.0 & 11.0 & 65.0 & 46.0 & 88.0 & 84.0 \\
& Adversarial Decoding
& 87.0 & 78.0 & 9.0 & 9.0 & 45.0 & 65.0 & 43.0 & 86.0 & 81.0 \\
& CPA-RAG
& 89.0 & 84.0 & 10.0 & 9.0 & 18.0 & 67.0 & 42.0 & 95.0 & 86.0 \\
\cmidrule(lr){2-11}
& PR-BB
& 89.0 & \textbf{84.0} & 2.0 & 2.0 & 4.0
& \textbf{72.0} & 50.0 & \textbf{91.0} & \textbf{87.0} \\
& \textbf{\method{}-BB (Ours)}
& \textbf{90.0} & \textbf{84.0} & \textbf{20.0} & \textbf{20.0}
& \textbf{57.0} & \textbf{72.0} & \textbf{61.0}
& \textbf{91.0} & \textbf{87.0} \\
\cmidrule(lr){2-11}
& PR-WB
& 90.0 & \textbf{82.0} & 2.0 & 2.0 & 9.0
& \textbf{70.0} & 46.0 & 88.0 & \textbf{87.0} \\
& \textbf{\method{}-WB (Ours)}
& \textbf{92.0} & 81.0 & \textbf{26.0} & \textbf{25.0}
& \textbf{68.0} & 69.0 & \textbf{61.0} & \textbf{89.0}
& \textbf{87.0} \\

\bottomrule
\end{tabular}%
}

\caption{
Attack performance across nine RAG configurations
(Vanilla plus eight defenses) on HotpotQA with poison budget \(B=5\)
and retrieval depth \(K=5\), measured by \textsc{strict-ASR}
(\%; higher is better for the attacker).
\textbf{PR} and \textbf{\method{}} denote PoisonedRAG and our method;
\textbf{BB} and \textbf{WB} denote their black-box and white-box access
settings.
The remaining rows are six cross-family reference baselines.
RobustRAG, Astute, and ReliabilityRAG denote RobustRAG-Keyword
(KeywordAgg), Astute-RAG, and ReliabilityRAG-MIS, respectively.
Within each matched PR--\method{} pair and access setting, bold marks the
higher value; exact ties are bolded for both.
}
\label{tab:attack-defense-hotpot-b5-k5}

\end{table*}

Averaged uniformly over the 27 target-generator--configuration
combinations, \method{} increases ASR over PR from \(53.37\%\) to
\(67.44\%\) under BB, a gain of \(14.07\) percentage points, and from
\(52.26\%\) to \(68.41\%\) under WB, a gain of \(16.15\) points.
Relative to PR, \method{} is higher in 19 BB cells, tied in six, and
lower in two; the corresponding WB counts are 21, four, and two.
After averaging over the nine configurations, the improvement remains
positive for each target LLM under both access settings.

\paragraph{Effectiveness under clustering-based filtering.}
Averaged over the three target LLMs, PR achieves \(2.33\%\) ASR
against TrustRAG under both BB and WB, whereas \method{} reaches
\(29.00\%\) under BB and \(36.33\%\) under WB.
The corresponding gains are \(26.67\) and \(34.00\) percentage points.
Against SeCon-RAG, \method{} raises ASR from \(2.33\%\) to \(26.33\%\)
under BB and from \(2.67\%\) to \(38.00\%\) under WB, yielding gains of
\(24.00\) and \(35.33\) points.

\method{} outperforms PR in all 12 matched comparisons spanning three
target LLMs, two clustering-based defenses, and both access settings.
The cell-level gains range from \(18\) to \(36\) points under BB and
from \(23\) to \(45\) points under WB.
Moreover, among \method{}-BB, PR-BB, and the six reference attacks,
\method{}-BB is uniquely highest in all six TrustRAG and SeCon-RAG cells.

These results show that \method{} remains substantially more effective
than PR under clustering-based filtering. As a separate output-level
diagnostic, Appendix~\ref{app:embedding-dispersion} shows that the final
\method{} bundles have lower mean and maximum within-bundle Contriever
cosine similarity than matched PR bundles across all four
dataset--access conditions, including BB without construction-time
access to Contriever. Section~5.3 examines how overlap control relates
to post-filter poison-slot survival.

\paragraph{Results on the remaining configurations.}
Averaged over the three target LLMs, \method{} improves ASR
against RAGuard over PR by \(58.00\) points under BB and \(60.00\)
points under WB.
Against Astute-RAG, the corresponding gains are \(15.00\) and \(8.67\)
points.
Differences under Vanilla, InstructRAG, RobustRAG-Keyword,
ReliabilityRAG-MIS, and RAGDefender are smaller on average.

Across \method{}-BB, PR-BB, and the six cross-family reference attacks,
\method{}-BB is uniquely highest in 16 cells and tied for highest in
eight, covering 24 of the 27 target-generator--configuration combinations.
Its average ASR of \(67.44\%\) exceeds that of the strongest reference
attack, Adversarial Decoding at \(55.22\%\), by \(12.22\) points.

\subsection{Ablation Studies}
\label{sec:ablations}

Table~\ref{tab:ablation-summary} reports independently reconstructed
ablations of Doubt allocation and the pairwise ROUGE-L
constraints, together with fixed-trajectory metric sensitivity.

\begin{table*}[t]
\centering
\small
\setlength{\tabcolsep}{3.2pt}
\renewcommand{\arraystretch}{1.10}

\begin{tabular*}{\textwidth}{
@{\extracolsep{\fill}}
c r r r
@{\hspace{14pt}}
l r r r
@{\hspace{14pt}}
l r r r
@{}
}
\toprule

\multicolumn{4}{c}{\textbf{(a) Doubt allocation}}
&
\multicolumn{4}{c}{\textbf{(b) ROUGE-L constraints}}
&
\multicolumn{4}{c}{\textbf{(c) Metric sensitivity}}
\\

\cmidrule(lr){1-4}
\cmidrule(lr){5-8}
\cmidrule(lr){9-12}

\(B\)
& \textbf{w/o Doubt}
& \textbf{Full}
& \(\mathbf{\Delta}\)
& \textbf{Constraints}
& \textbf{Van.}
& \textbf{Def.}
& \textbf{Surv.}
& \textbf{Metric}
& \textbf{Van.}
& \textbf{Def.}
& \textbf{Surv.}
\\
\midrule

2
& 45.5
& \textbf{52.8}
& \textbf{+7.2}
& \textbf{Max + Mean}
& 93.0
& 43.6
& \textbf{90.5}
& \textbf{ROUGE-L}
& 93.0
& \textbf{43.6}
& \textbf{90.5}
\\

3
& 51.7
& \textbf{58.8}
& \textbf{+7.1}
& Max only
& 93.0
& \textbf{43.8}
& 90.3
& Token Jaccard
& \textbf{94.5}
& 38.0
& 64.9
\\

4
& 53.9
& \textbf{60.4}
& \textbf{+6.4}
& Mean only
& 93.2
& 40.9
& 81.0
& Char. TF--IDF
& 93.2
& 36.9
& 73.5
\\

5
& 57.4
& \textbf{64.7}
& \textbf{+7.3}
& Neither
& \textbf{94.8}
& 26.5
& 42.9
& Sup. SimCSE
& 93.2
& 40.8
& 84.9
\\

\bottomrule
\end{tabular*}

\par\vspace{3pt}

\begin{minipage}{\textwidth}
\scriptsize
\raggedright
\textit{Scope.}
Panel (a) macro-averages over nine RAG configurations, two datasets,
three target LLMs, and both access settings; 
Full uses one Doubt passage and \(B-1\) direct-support passages, whereas
\emph{w/o Doubt} uses \(B\) direct-support passages.
Panels (b)--(c) use Qwen2.5-7B on HotpotQA and NQ under both access settings.
Van. denotes Vanilla ASR; Def. averages TrustRAG and SeCon-RAG.
Surv. denotes the macro-averaged percentage of retrieved poison slots
retained after the \(k\)-means/\(n\)-gram pre-filter used in our
TrustRAG and SeCon-RAG implementations.
Results are compared within panels.
\end{minipage}

\par

\caption{
Construction ablations and overlap-metric sensitivity.
ASR and survival are percentages; \(\Delta\) is in percentage points
and is computed before rounding.
Panel (a) evaluates the allocation of one poison slot to Doubt.
Panel (b) ablates the maximum and mean ROUGE-L constraints.
Panel (c) compares overlap metrics on shared ROUGE-L-guided candidate
and repair trajectories while retaining both constraints.
Higher is better for the attacker.
}
\label{tab:ablation-summary}

\end{table*}

\paragraph{Contribution of Doubt.}
Across \(B\in\{2,3,4,5\}\), Full---one Doubt passage and \(B-1\)
direct-support passages---outperforms the all-direct-support allocation
at every budget by \(6.4\)--\(7.3\) percentage points
(Table~\ref{tab:ablation-summary}(a)).
Because each arm independently reconstructs the complete poison set,
this comparison evaluates the intended set-level budget-allocation policy,
including the interaction between Doubt and direct support.

\paragraph{Pairwise lexical-overlap control.}
Removing both ROUGE-L constraints reduces defended ASR from \(43.6\%\)
to \(26.5\%\) and post-filter survival from \(90.5\%\) to \(42.9\%\).
By contrast, Vanilla ASR remains between \(93.0\%\) and \(94.8\%\)
across all four configurations.
This separation links the benefit of overlap control to poison-slot
survival under filtering rather than unfiltered attack strength.

The two constraints contribute asymmetrically.
At the point-estimate level, Max-only nearly matches Max+Mean in
defended ASR (\(43.8\%\) vs.\ \(43.6\%\)) and survival
(\(90.3\%\) vs.\ \(90.5\%\)).
Mean-only remains substantially stronger than Neither in defended ASR
(\(40.9\%\) vs.\ \(26.5\%\)) and survival
(\(81.0\%\) vs.\ \(42.9\%\)).
Thus, the maximum constraint is the stronger individual control, while
the mean constraint remains beneficial on its own.

\paragraph{Sensitivity to the overlap metric.}
Within this fixed-trajectory comparison, ROUGE-L yields the highest
defended ASR and post-filter survival, at \(43.6\%\) and \(90.5\%\),
respectively.
Token Jaccard, character TF--IDF, and supervised SimCSE
\citep{gao-etal-2021-simcse} yield defended
ASR values of \(38.0\%\), \(36.9\%\), and \(40.8\%\), with survival
rates of \(64.9\%\), \(73.5\%\), and \(84.9\%\).
Vanilla ASR remains high for every metric
(\(93.0\%\)--\(94.5\%\)); the metric-dependent performance separation
is therefore concentrated in the filtering.

\section{Conclusion}
In this paper, we introduce \method{}, a distributed corpus-poisoning attack
designed for defenses that jointly compare retrieved evidence. \method{}
constructs passages with distinct document styles and target-support framings,
controls pairwise lexical overlap, and reserves one passage to cast doubt on
evidence for the reference answer. Together, these passages preserve agreement
on the attacker-chosen conclusion while dispersing their representations and
wording, weakening similarity- and conflict-aware defenses. Evaluations on
HotpotQA and Natural Questions with three target LLMs and nine RAG settings
show that \method{} matches or outperforms prior attacks in most
configurations. Our results indicate that filtering for semantic concentration
alone is insufficient and motivate defenses that can identify coordinated
evidence manipulation across stylistically distinct passages.

\section*{Limitations}
Our evaluation is limited in scale by available computational and API
budgets. We conduct experiments on fixed 100-question subsets of
HotpotQA and Natural Questions and evaluate three open-weight target
generators; proprietary closed-source models are not included. Although
we cover a representative set of RAG defenses, the rapidly evolving
defense landscape means that some recently proposed defenses are not
evaluated. Future work could extend the evaluation to larger query sets,
additional datasets, closed-source models, and newer defenses to further
assess the generalizability of our findings.

\section*{Ethical Considerations}

This work studies a dual-use security problem. The proposed techniques
could be misused to manipulate answers produced by RAG systems whose
knowledge bases accept untrusted content. Our purpose is to evaluate the
robustness of existing defenses under controlled conditions and to
motivate stronger protections against coordinated corpus poisoning.

All experiments were conducted offline using public research
benchmarks and locally instantiated RAG pipelines. We did not attack
deployed services, modify third-party corpora, publish poisoned passages
to the Web, recruit human participants, or collect new personal data.
The generated passages were used only within the benchmark evaluation
environment.

\bibliographystyle{acl_natbib}
\bibliography{custom}

\appendix
\section{Additional Evaluation Results}

\label{app:additional-results}

This appendix extends the main HotpotQA evaluation in four directions.
We first report the complete Natural Questions (NQ)
attack--generator--configuration matrix under the default setting,
then examine sensitivity to the dense retriever, evaluate
poison-budget sensitivity on HotpotQA, and finally report auxiliary
target-answer precision and \(F_1\).

Throughout this appendix, \textbf{BB} and \textbf{WB} denote black-box and white-box retriever access, respectively.

\subsection{Results on Natural Questions}
\label{app:nq-results}

We test whether the aggregate HotpotQA advantage transfers to NQ under the
same default poison budget and retrieval depth (\(B=K=5\)).
Table~\ref{tab:attack-defense-nq-b5-k5} reports the complete matrix for three
target LLMs and nine RAG configurations (Vanilla plus eight defenses).
We use the same 100-question evaluation size, attack definitions,
retriever-access settings, and \textsc{strict-ASR} metric as in the main
evaluation.

\begin{table*}[t]
\centering
\scriptsize
\setlength{\tabcolsep}{2.2pt}
\renewcommand{\arraystretch}{1.04}

\resizebox{\textwidth}{!}{%
\begin{tabular}{@{}ll|*{9}{c}@{}}
\toprule

\multirow{2}{*}{\textbf{Target LLM}} &
\multirow{2}{*}{\textbf{Attack}} &
\multicolumn{9}{c}{\textbf{RAG configuration}} \\
\cmidrule(lr){3-11}
& &
\textbf{Vanilla} &
\textbf{Instruct} &
\textbf{TrustRAG} &
\textbf{SeCon} &
\textbf{RAGuard} &
\textbf{RobustRAG} &
\textbf{Astute} &
\textbf{ReliabilityRAG} &
\textbf{RAGDefender} \\
\midrule

\multirow{10}{*}{\textbf{Llama}}
& CorpusPoisoning
& 0.0 & 0.0 & 2.0 & 2.0 & 0.0 & 2.0 & 0.0 & 0.0 & 0.0 \\
& PromptInjection
& 40.0 & 25.0 & 4.0 & 3.0 & 0.0 & 12.0 & 18.0 & 17.0 & 0.0 \\
& Paradox
& 39.0 & 32.0 & 5.0 & 4.0 & 33.0 & 20.0 & 7.0 & 34.0 & 31.0 \\
& Disinformation
& 38.0 & 35.0 & 3.0 & 4.0 & 36.0 & 19.0 & 15.0 & 28.0 & 30.0 \\
& Adversarial Decoding
& 72.0 & 69.0 & 3.0 & 3.0 & 64.0 & 43.0 & 30.0 & 80.0 & 60.0 \\
& CPA-RAG
& 87.0 & 74.0 & 4.0 & 3.0 & 70.0 & 46.0 & 19.0 & 82.0 & 67.0 \\
\cmidrule(lr){2-11}
& PR-BB
& \textbf{86.0} & 68.0 & 3.0 & 3.0 & 53.0
& \textbf{38.0} & 20.0 & 88.0 & 75.0 \\
& \textbf{\method{}-BB (Ours)}
& 85.0 & \textbf{77.0} & \textbf{14.0} & \textbf{15.0}
& \textbf{78.0} & 32.0 & \textbf{30.0} & \textbf{90.0}
& \textbf{78.0} \\
\cmidrule(lr){2-11}
& PR-WB
& \textbf{95.0} & 72.0 & 2.0 & 5.0 & 22.0
& \textbf{40.0} & 26.0 & \textbf{93.0} & 79.0 \\
& \textbf{\method{}-WB (Ours)}
& 92.0 & \textbf{84.0} & \textbf{19.0} & \textbf{20.0}
& \textbf{82.0} & 39.0 & \textbf{36.0} & \textbf{93.0}
& \textbf{88.0} \\
\midrule

\multirow{10}{*}{\textbf{Qwen}}
& CorpusPoisoning
& 3.0 & 4.0 & 3.0 & 3.0 & 4.0 & 0.0 & 3.0 & 0.0 & 2.0 \\
& PromptInjection
& 65.0 & 41.0 & 11.0 & 8.0 & 4.0 & 19.0 & 22.0 & 21.0 & 4.0 \\
& Paradox
& 44.0 & 39.0 & 8.0 & 8.0 & 40.0 & 23.0 & 17.0 & 32.0 & 33.0 \\
& Disinformation
& 40.0 & 37.0 & 4.0 & 4.0 & 40.0 & 18.0 & 15.0 & 32.0 & 31.0 \\
& Adversarial Decoding
& 82.0 & 76.0 & 3.0 & 3.0 & 77.0 & 62.0 & 39.0 & 88.0 & 84.0 \\
& CPA-RAG
& 92.0 & 75.0 & 6.0 & 7.0 & 82.0 & 59.0 & 42.0 & 92.0 & 88.0 \\
\cmidrule(lr){2-11}
& PR-BB
& \textbf{92.0} & 71.0 & 4.0 & 4.0 & 62.0
& \textbf{63.0} & 41.0 & \textbf{92.0} & 86.0 \\
& \textbf{\method{}-BB (Ours)}
& 89.0 & \textbf{85.0} & \textbf{23.0} & \textbf{22.0}
& \textbf{83.0} & 60.0 & \textbf{44.0} & 90.0
& \textbf{89.0} \\
\cmidrule(lr){2-11}
& PR-WB
& \textbf{94.0} & 72.0 & 4.0 & 7.0 & 27.0
& \textbf{63.0} & \textbf{41.0} & \textbf{97.0}
& \textbf{92.0} \\
& \textbf{\method{}-WB (Ours)}
& \textbf{94.0} & \textbf{91.0} & \textbf{37.0} & \textbf{36.0}
& \textbf{85.0} & \textbf{63.0} & 33.0 & 94.0 & 89.0 \\
\midrule

\multirow{10}{*}{\textbf{Mistral}}
& CorpusPoisoning
& 3.0 & 4.0 & 1.0 & 1.0 & 3.0 & 5.0 & 2.0 & 1.0 & 3.0 \\
& PromptInjection
& 43.0 & 29.0 & 7.0 & 9.0 & 3.0 & 20.0 & 14.0 & 16.0 & 3.0 \\
& Paradox
& 42.0 & 43.0 & 1.0 & 1.0 & 39.0 & 18.0 & 11.0 & 33.0 & 35.0 \\
& Disinformation
& 36.0 & 37.0 & 2.0 & 2.0 & 37.0 & 17.0 & 13.0 & 31.0 & 33.0 \\
& Adversarial Decoding
& 85.0 & 81.0 & 3.0 & 2.0 & 78.0 & 50.0 & 33.0 & 89.0 & 83.0 \\
& CPA-RAG
& 91.0 & 87.0 & 3.0 & 3.0 & 77.0 & 48.0 & 27.0 & 87.0 & 87.0 \\
\cmidrule(lr){2-11}
& PR-BB
& \textbf{91.0} & 88.0 & 5.0 & 4.0 & 60.0
& \textbf{48.0} & 36.0 & \textbf{95.0} & \textbf{85.0} \\
& \textbf{\method{}-BB (Ours)}
& 89.0 & \textbf{89.0} & \textbf{18.0} & \textbf{19.0}
& \textbf{82.0} & 42.0 & \textbf{39.0} & 92.0 & 83.0 \\
\cmidrule(lr){2-11}
& PR-WB
& \textbf{96.0} & 89.0 & 4.0 & 6.0 & 21.0
& \textbf{50.0} & 32.0 & 94.0 & \textbf{89.0} \\
& \textbf{\method{}-WB (Ours)}
& 94.0 & \textbf{93.0} & \textbf{24.0} & \textbf{26.0}
& \textbf{88.0} & \textbf{50.0} & \textbf{43.0}
& \textbf{95.0} & \textbf{89.0} \\

\bottomrule
\end{tabular}%
}

\caption{
Attack performance across nine RAG configurations
(Vanilla plus eight defenses) on Natural Questions (NQ) with poison budget
\(B=5\) and retrieval depth \(K=5\), measured by \textsc{strict-ASR}
(\%; higher is better for the attacker).
\textbf{PR} and \textbf{\method{}} denote PoisonedRAG and our method;
\textbf{BB} and \textbf{WB} denote their black-box and white-box access
settings.
The remaining rows are six cross-family reference baselines.
RobustRAG, Astute, and ReliabilityRAG denote RobustRAG-Keyword
(KeywordAgg), Astute-RAG, and ReliabilityRAG-MIS, respectively;
RAGuard denotes RAGuard.
Within each matched PR--\method{} pair and access setting, bold marks the
higher value; exact ties are bolded for both.
}
\label{tab:attack-defense-nq-b5-k5}

\end{table*}

Across the 27 matched target-generator--configuration cells, \method{}
increases \textsc{strict-ASR} over PR from \(54.11\%\) to \(60.63\%\)
under BB, a gain of \(6.52\) percentage points.
Under WB, the corresponding average increases from \(52.30\%\) to
\(65.81\%\), a gain of \(13.52\) points.
Gains are computed before rounding the displayed averages.
Relative to PR, \method{} is higher/tied/lower in \(18/0/9\) cells under BB
and \(16/5/6\) cells under WB.
After averaging over the nine configurations separately for each target
LLM, \method{} improves over PR for all three generators under both
access settings.

The gains are especially consistent for InstructRAG, TrustRAG, SeCon-RAG,
and RAGuard: for each configuration, \method{} improves over the
access-matched PR baseline in all six generator--access comparisons.
Results for the remaining configurations are mixed.
Thus, NQ supports transfer of the aggregate advantage beyond HotpotQA, not
uniform cell-wise dominance.

\subsection{Sensitivity to the Dense Retriever}
\label{app:retriever-sensitivity}

\paragraph{Setup.}
We assess the sensitivity of \method{}-BB to the choice of dense retriever at
\(B=K=5\).
Within each dataset, this auxiliary run holds the evaluation questions,
constructed poison passages, target LLMs, and downstream configurations
fixed while replacing Contriever with DPR-single
\citep{karpukhin-etal-2020-dense} or ANCE
\citep{xiong-etal-2021-ance}.
The experiment contains eight RAG configurations: it includes Perplexity and
excludes ReliabilityRAG-MIS and RAGDefender.
Because this run is separate from the main nine-configuration evaluation, its
Contriever results serve only as the within-run reference and are not intended
to reproduce the main matrix.

Table~\ref{tab:retriever-summary} averages \textsc{strict-ASR} uniformly over
HotpotQA, NQ, and the three target LLMs; its mean row additionally
averages over the eight configurations.
For a dataset with \(N=100\) questions, the poison retrieval recall is
\(100R/(BN)\), where \(R\) is the total number of query-specific poison
passages appearing in their corresponding undefended top-\(K\) contexts.
\begin{table}[t]
\centering

{\scriptsize
\setlength{\tabcolsep}{3.2pt}

\begin{tabular}{@{}lrrr@{}}
\toprule
\textbf{Configuration}
& \textbf{Contriever}
& \textbf{DPR-single}
& \textbf{ANCE} \\
\midrule
Vanilla     & 85.5 & 61.8 & 84.5 \\
InstructRAG & 80.0 & 54.8 & 76.7 \\
TrustRAG    & 23.7 & 13.0 & 21.5 \\
SeCon-RAG   & 22.3 & 14.3 & 22.5 \\
Perplexity  & 85.2 & 61.2 & 84.7 \\
RAGuard     & 69.5 & 48.2 & 70.5 \\
RobustRAG   & 56.5 & 39.2 & 59.2 \\
Astute-RAG  & 48.7 & 33.7 & 52.3 \\
\midrule
\textbf{Mean}
& \textbf{58.9}
& \textbf{40.8}
& \textbf{59.0} \\
\emph{Poison retrieval recall}
& 85.9
& 42.6
& 81.1 \\
\bottomrule
\end{tabular}
}

\par

\caption{
Dense-retriever sensitivity of \method{}-BB at \(B=K=5\).
Each \textsc{strict-ASR} entry (\%) is averaged over both datasets and all
three target LLMs, and the mean row assigns equal weight to the eight
RAG configurations.
The final row reports the poison retrieval recall (\%) before any defense,
averaged over both datasets.
}
\label{tab:retriever-summary}

\end{table}

\paragraph{Results.}
ANCE yields essentially the same mean \textsc{strict-ASR} as Contriever
(\(59.0\%\) versus \(58.9\%\)), whereas the value with DPR-single is
\(40.8\%\).
DPR-single also has a substantially lower poison retrieval recall
(\(42.6\%\), versus \(85.9\%\) for Contriever and \(81.1\%\) for ANCE).
The lower ASR therefore co-occurs with lower poison retrieval recall.
This pattern is consistent with poison retrieval recall contributing to the
difference, but does not isolate that factor, because replacing the retriever
also changes passage rankings and the complete retrieved context.
Table~\ref{tab:retriever-hp-full} reports the complete HotpotQA breakdown.
\begin{table*}[t]
\centering

{\scriptsize
\setlength{\tabcolsep}{3.4pt}
\renewcommand{\arraystretch}{1.04}

\resizebox{\textwidth}{!}{%
\begin{tabular}{@{}lrrr rrr rrr@{}}
\toprule
& \multicolumn{3}{c}{\textbf{Contriever}}
& \multicolumn{3}{c}{\textbf{DPR-single}}
& \multicolumn{3}{c}{\textbf{ANCE}} \\
\cmidrule(lr){2-4}
\cmidrule(lr){5-7}
\cmidrule(lr){8-10}

\textbf{RAG configuration}
& \textbf{Llama}
& \textbf{Qwen}
& \textbf{Mistral}
& \textbf{Llama}
& \textbf{Qwen}
& \textbf{Mistral}
& \textbf{Llama}
& \textbf{Qwen}
& \textbf{Mistral} \\
\midrule

Vanilla
& 86 & 91 & 88
& 74 & 74 & 73
& 84 & 88 & 86 \\

InstructRAG
& 79 & 80 & 83
& 60 & 66 & 70
& 73 & 78 & 78 \\

TrustRAG
& 41 & 29 & 21
& 20 & 23 & 17
& 22 & 29 & 19 \\

SeCon-RAG
& 34 & 31 & 17
& 27 & 25 & 18
& 24 & 28 & 20 \\

Perplexity
& 85 & 91 & 87
& 73 & 73 & 72
& 84 & 88 & 86 \\

RAGuard
& 61 & 65 & 55
& 52 & 49 & 52
& 61 & 63 & 59 \\

RobustRAG-Keyword
& 58 & 76 & 72
& 44 & 61 & 56
& 62 & 76 & 70 \\

Astute-RAG
& 60 & 59 & 64
& 51 & 42 & 46
& 72 & 55 & 57 \\

\midrule

\emph{Poison retrieval recall}
& \multicolumn{3}{c}{98.4}
& \multicolumn{3}{c}{52.2}
& \multicolumn{3}{c}{86.4} \\

\bottomrule
\end{tabular}%
}
}

\par

\caption{
Complete dense-retriever sensitivity results for \method{}-BB on HotpotQA at
\(B=K=5\), measured by \textsc{strict-ASR} (\%).
Each generator entry is computed over the same 100 questions.
The poison retrieval recall is computed from the undefended retrieval results
and is shared across target LLMs; it is therefore reported once per
retriever.
}
\label{tab:retriever-hp-full}

\end{table*}

\subsection{Poison-Budget Performance on HotpotQA}
\label{app:hp-budget}

\begin{figure*}[t]
\centering
\includegraphics[width=0.82\textwidth]
{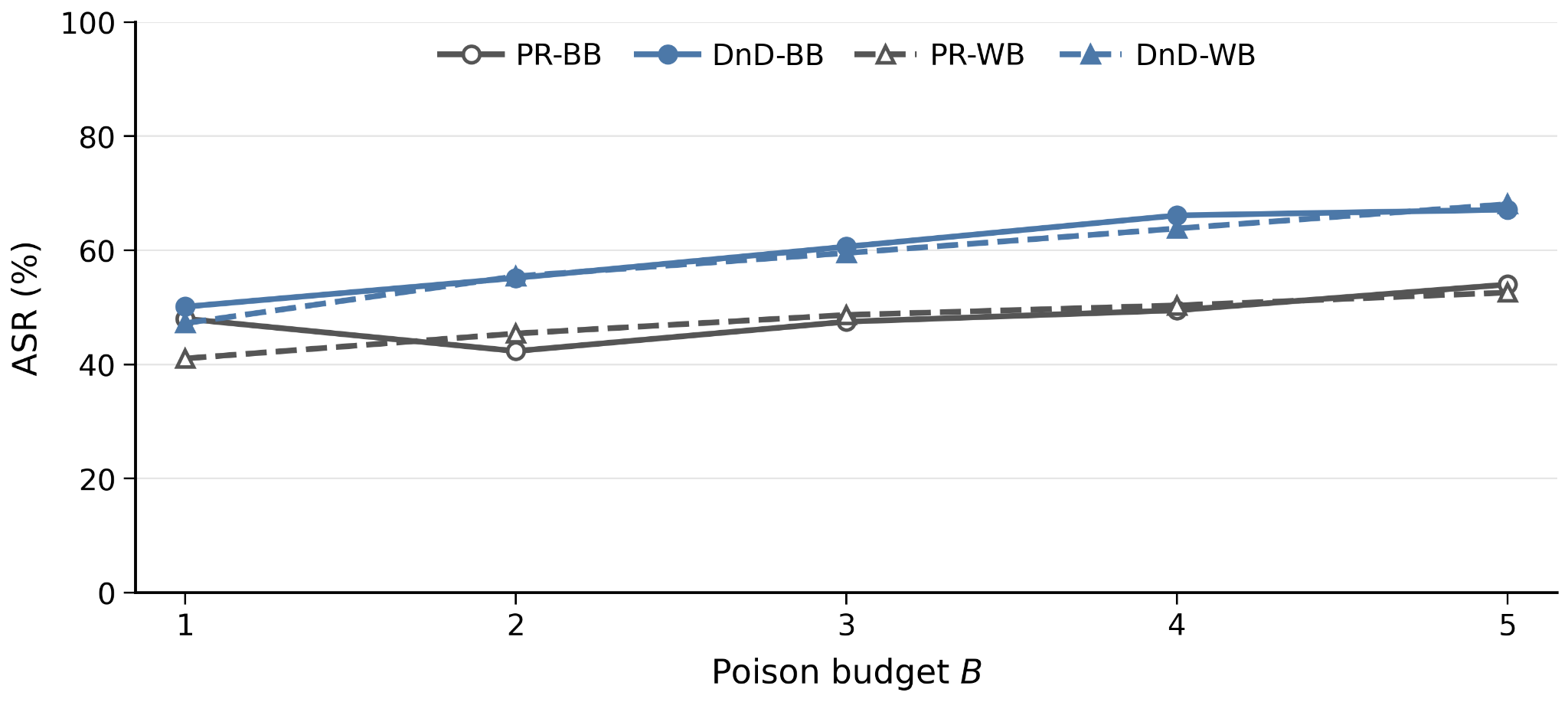}
\caption{
Poison-budget performance on HotpotQA at fixed retrieval depth \(K=5\).
The single-passage setting uses direct support without Doubt, whereas the
multi-passage settings combine direct support with a Doubt passage.
Each point reports \textsc{strict-ASR} over three target LLMs and
nine RAG configurations (27 cells).
PR and \method{} are evaluated under matched black-box (BB) and white-box (WB)
retriever access.
Higher is better for the attacker.
}
\label{fig:budget-scaling-hp}
\end{figure*}

We vary the poison budget from \(B=1\) to \(B=5\) on HotpotQA while fixing
the retrieval depth at \(K=5\).
The single-passage setting uses one direct-support passage without Doubt, whereas the multi-passage settings combine direct support with a Doubt passage.
Figure~\ref{fig:budget-scaling-hp} reports PR and \method{} under matched BB and WB retriever access.
Each point is the average over three target LLMs and the same nine
RAG configurations used in the main HotpotQA evaluation (27 cells).

In these 27-cell averages, \method{} exceeds the access-matched PR
baseline at all five budgets under both access settings.
The \method{}-BB and \method{}-WB averages increase monotonically from
\(B=1\) to \(B=5\), whereas PR-BB decreases from \(B=1\) to \(B=2\) before
increasing from \(B=2\) to \(B=5\).
Thus, the aggregate advantage is present in the single-passage case and
persists throughout the evaluated multi-passage range.

\subsection{Auxiliary Metrics}
\label{app:auxiliary-metrics}

To provide a finer-grained evaluation of target-answer realization, we
additionally report target-answer precision and \(F_1\).
We normalize the generated response and target answer \(y^\star\) by
lowercasing, removing punctuation and the English articles
\textit{a}, \textit{an}, and \textit{the}, and normalizing whitespace.
We then compute their multiset token overlap.
Precision measures the proportion of normalized response tokens matched
by \(y^\star\), while \(F_1\) jointly captures precision and target-answer
coverage.

Both metrics are computed separately for each response and then equally
averaged over the three target LLMs and nine RAG configurations
within each dataset.

\begin{table}[t]
\centering
\small
\setlength{\tabcolsep}{3.5pt}
\renewcommand{\arraystretch}{1.08}

\begin{tabular}{@{}lrrrr@{}}
\toprule
\multirow{2}{*}{\textbf{Attack}} &
\multicolumn{2}{c}{\textbf{HotpotQA}} &
\multicolumn{2}{c}{\textbf{NQ}} \\
\cmidrule(lr){2-3}
\cmidrule(lr){4-5}
&
\textbf{Prec.} &
\textbf{\(F_1\)} &
\textbf{Prec.} &
\textbf{\(F_1\)} \\
\midrule

CorpusPoisoning
& 3.0 & 4.1
& 1.9 & 2.5 \\

Disinformation
& 25.7 & 29.7
& 10.4 & 12.4 \\

Paradox
& 25.0 & 29.0
& 9.6 & 11.7 \\

Adversarial Decoding
& 29.3 & 32.7
& 23.4 & 27.6 \\

CPA-RAG
& 29.6 & 33.0
& \underline{27.5} & \underline{31.5} \\

\midrule

PR-BB
& 27.4 & 31.0
& 25.8 & 30.0 \\

\textbf{\method{}-BB (Ours)}
& \textbf{33.7} & \textbf{37.5}
& 24.2 & 28.3 \\

\midrule

PR-WB
& 25.8 & 29.6
& 22.5 & 26.6 \\

\textbf{\method{}-WB (Ours)}
& \underline{31.6} & \underline{35.5}
& \textbf{28.2} & \textbf{32.7} \\

\bottomrule
\end{tabular}

\caption{
Auxiliary target-answer precision and \(F_1\) scores
(\%; higher is better) under the default poison budget \(B=5\)
and retrieval depth \(K=5\).
Each value is averaged over the \(3\times9=27\)
target-generator--RAG-configuration cells for the corresponding dataset.
The BB/WB suffixes denote the retriever-access setting for PR and
\method{}.
Bold and underline indicate the highest and second-highest result in
each column, respectively.
}
\label{tab:auxiliary-metrics}
\end{table}

\paragraph{Results.}
Table~\ref{tab:auxiliary-metrics} shows that a \method{} variant achieves
the highest precision and \(F_1\) on each benchmark.
On HotpotQA, \method{}-BB ranks first on both metrics, while
\method{}-WB ranks second.
Relative to access-matched PR, \method{} improves precision and \(F_1\)
by 6.3 and 6.5 percentage points under BB access and by 5.8 and
5.9 points under WB access.

On NQ, \method{}-WB achieves the highest precision and \(F_1\),
outperforming PR-WB by 5.7 and 6.1 percentage points, respectively.
These results provide additional evidence that \method{} effectively
realizes the target answer, with particularly consistent improvements
under WB access across both benchmarks.

\subsection{Representation-Space Diversity of Final Poison Bundles}
\label{app:embedding-dispersion}

We complement the lexical-overlap analysis with a
representation-space diagnostic of the final poison bundles.
ROUGE-L and Contriever cosine similarity characterize complementary
properties: the former measures lexical sequence overlap, whereas the
latter measures concentration in the evaluated retriever's learned
representation space. We therefore compare the final \method{} and PR
bundles directly under the matched conditions described in
Section~5.1.

\paragraph{Metric.}
For each question, we use the exact \(B=5\) passages that enter the
end-to-end retrieval pipeline. Because each constructed passage set is
frozen and reused across the three target LLMs, every unique
question-level bundle is encoded once. We encode each passage with
\texttt{facebook/contriever}, using average pooling and the same
512-token input limit as the retrieval pipeline. The pooled
representations are \(\ell_2\)-normalized for cosine computation.
All evaluated passages fall within the input limit, so no passage is
truncated.

Let \(z_1,\ldots,z_B\) denote the normalized embeddings of the
passages in one bundle. We compute
\begin{equation}
S_{\mathrm{mean}}
=
\frac{1}{\binom{B}{2}}
\sum_{i<j} z_i^\top z_j,
\qquad
S_{\max}
=
\max_{i<j} z_i^\top z_j .
\label{eq:embedding-concentration}
\end{equation}
For \(B=5\), both statistics are computed over the ten unordered
passage pairs. \(S_{\mathrm{mean}}\) measures aggregate
within-bundle concentration, while \(S_{\max}\) captures the most
similar passage pair. Lower values indicate greater
representation-space dispersion.

All comparisons are paired by question ID. Confidence intervals use
20,000 paired bootstrap samples, and two-sided \(p\)-values use
100,000 paired sign-flip randomizations.

\begin{table*}[t]
\centering
{
\scriptsize
\setlength{\tabcolsep}{2.8pt}
\renewcommand{\arraystretch}{1.08}
\begin{tabular}{@{}llrrc@{\hspace{7pt}}rrc@{}}
\toprule
& &
\multicolumn{3}{c}{\(S_{\mathrm{mean}}\downarrow\)} &
\multicolumn{3}{c}{\(S_{\max}\downarrow\)} \\
\cmidrule(lr){3-5}
\cmidrule(lr){6-8}
\textbf{Dataset}
& \textbf{Access}
& \textbf{\method{}}
& \textbf{PR}
& \(\Delta\) \textbf{[95\% CI]}
& \textbf{\method{}}
& \textbf{PR}
& \(\Delta\) \textbf{[95\% CI]} \\
\midrule
HotpotQA & BB
& \textbf{0.7089}
& 0.8972
& \(-0.1884\,[-0.1995,-0.1774]\)
& \textbf{0.7684}
& 0.9351
& \(-0.1668\,[-0.1775,-0.1563]\) \\

HotpotQA & WB
& \textbf{0.6610}
& 0.7031
& \(-0.0421\,[-0.0493,-0.0350]\)
& \textbf{0.7169}
& 0.7651
& \(-0.0482\,[-0.0570,-0.0394]\) \\

NQ & BB
& \textbf{0.6335}
& 0.8457
& \(-0.2122\,[-0.2206,-0.2038]\)
& \textbf{0.7349}
& 0.8997
& \(-0.1647\,[-0.1742,-0.1553]\) \\

NQ & WB
& \textbf{0.6075}
& 0.6555
& \(-0.0480\,[-0.0567,-0.0394]\)
& \textbf{0.6760}
& 0.7311
& \(-0.0551\,[-0.0658,-0.0447]\) \\
\bottomrule
\end{tabular}
}
\caption{
Within-bundle Contriever representation concentration for the exact
final \(B=5\) poison bundles. Lower values indicate greater
representation-space dispersion. Differences are computed as
\(\method{}-\mathrm{PR}\) from paired, unrounded question-level
values; brackets report paired 95\% bootstrap confidence intervals.
}
\label{tab:embedding-dispersion}
\end{table*}

\paragraph{Final-bundle comparison.}
Table~\ref{tab:embedding-dispersion} shows a consistent
representation-space advantage for \method{}. Across all four
dataset--access conditions, \method{} achieves lower
\(S_{\mathrm{mean}}\) and \(S_{\max}\) than matched PR, with every
paired 95\% confidence interval lying below zero
(\(p<0.0001\) in all comparisons).

The differences are particularly pronounced under BB access.
On HotpotQA and NQ, respectively, \method{} reduces
\(S_{\mathrm{mean}}\) by \(0.1884\) and \(0.2122\), and reduces
\(S_{\max}\) by \(0.1668\) and \(0.1647\). Moreover, every BB
question has lower \(S_{\mathrm{mean}}\) and \(S_{\max}\) under
\method{} than under PR. This result is obtained without either
attack receiving rankings, scores, parameters, or gradients from
the evaluated Contriever during construction.

Under WB access, where both attacks use the same Contriever access
and HotFlip optimization budget, the differences remain consistent
across both datasets. \method{} has lower \(S_{\mathrm{mean}}\) for
88\% of questions on each dataset and lower \(S_{\max}\) for 83\%
of HotpotQA questions and 89\% of NQ questions. The paired
confidence intervals remain separated from zero for both statistics.

\paragraph{Pre-adaptation representation structure.}
The representation-space difference is already present in the frozen
passage bodies before access-specific retrieval adaptation. Relative
to PR, \method{} reduces body-level \(S_{\mathrm{mean}}\) by
\(0.1424\) on HotpotQA and \(0.1353\) on NQ. The corresponding
reductions in body-level \(S_{\max}\) are \(0.1428\) and \(0.1397\).
The difference then remains visible after both BB and WB retrieval
adaptation. Thus, the distinct representation structure is established
during passage construction and preserved through the retrieval-facing
stage.

Overall, the final \method{} bundles are less concentrated than
matched PR bundles in the evaluated Contriever representation space
across every dataset and retriever-access condition. These findings
provide direct representation-level evidence that the diversity of
\method{} extends beyond lexical overlap and is preserved in the
passages ultimately used by the RAG pipeline.

\section{Implementation Details}
\label{app:implementation}

\paragraph{Data Selection and Evaluation Protocol.}
We shuffle each benchmark using \texttt{seed=12} and traverse the resulting
fixed order. We retain the first 100 non-yes/no questions for which the
reference answer \(y\) is unambiguous and a distinct, type-compatible target
answer \(y^\star\) can be registered. Eligibility is determined before attack
construction and does not use retrieval results, target-model outputs, defense
outcomes, or \textsc{strict-ASR}. The resulting query--target sets are then
frozen and shared across all attacks, target LLMs, and RAG configurations.

Our default evaluation uses poison budget \(B=5\), retrieval depth \(K=5\),
three target LLMs, and nine RAG configurations (Vanilla plus eight defenses).
Each reported cell evaluates the corresponding fixed 100-query set using
\texttt{seed=12}. PoisonedRAG and \method{} use identical query--target pairs,
corpora, poison budgets, retrievers, target models, and RAG configurations.

\paragraph{Models.}
For both PR and \method{}, we use gpt-5-mini as the
construction-time writer and \nolinkurl{google/gemma-4-E4B-it}  as the construction-time verifier. The three target LLMs are \nolinkurl{meta-llama/Meta-Llama-3.1-8B-Instruct}, \texttt{Qwen/Qwen2.5-7B-Instruct}, and \texttt{mistralai/Mistral-7B-Instruct-v0.3}. The default dense retriever is \texttt{facebook/contriever}. Gemma is used only to verify construction-time constraints. It returns structured passage- and bundle-level validation results but has no access to retrieval results, target-model outputs, defense outcomes, or ASR.

\paragraph{Attack Construction.}
At \(B=5\), each \method{} bundle contains four complementary direct-support
passages and one \emph{Doubt} passage. For each query, the writer may initiate
at most three construction trajectories. Within each trajectory, we permit at
most five accepted local repairs, with no more than two failed passages modified
during each repair. We freeze the first bundle that satisfies all construction
checks; no retrieval- or ASR-based candidate selection is performed.

Gemma uses greedy decoding with \texttt{do\_sample=False}, disabled thinking, and \texttt{max\_new\_tokens=128}. We measure passage diversity using pairwise ROUGE-L F1 from \texttt{rouge-score==0.1.2}, with stemming enabled, over all ten unordered passage pairs. For the frozen semantic bodies, the maximum and mean pairwise scores must not exceed \(0.25\) and \(0.220\), respectively. After BB-side formatting, the corresponding limits are \(0.25\) and \(0.235\). The ROUGE-L thresholds were fixed before the reported evaluation and were not selected using retrieval results, target-LLM outputs, defense outcomes, or ASR on either evaluation set.

\paragraph{Retriever Access and Optimization.}
Under BB access, attack construction does not use retriever gradients. Under WB
access, we apply HotFlip independently to each passage, initialized from the
original question. We use 30 iterations, 100 candidates per iteration,
dot-product optimization, and \texttt{seed=12}. The semantic passage bodies
remain frozen throughout retrieval optimization.

Contriever uses average pooling without L2 normalization, dot-product
similarity, and exact retrieval.

\paragraph{Target Generation.}
We run the target LLMs using the \texttt{lmdeploy}
\texttt{PytorchEngine} in FP16. We pass a temperature of \(0.01\),
\texttt{max\_new\_tokens=500}, batch size \(1\), and a maximum prefill length
of \(8192\).
We report the supplied generation parameters without characterizing decoding
as greedy or sampling, because this behavior depends on the LMDeploy engine
configuration.

\paragraph{RAG Configurations.}
The nine evaluated RAG configurations are Vanilla (no defense), InstructRAG,
TrustRAG, SeCon-RAG, RobustRAG-Keyword, ReliabilityRAG-MIS,
RAGDefender, RAGuard, and Astute-RAG. We follow the official implementations
and their reported settings for all defenses, and use identical configurations
for every attack.

Following these official configurations, TrustRAG and SeCon-RAG use a threshold
of \(0.88\); RobustRAG uses \(\alpha=0.3\) and \(\beta=3\);
RAGDefender uses \texttt{multihop} mode for HotpotQA and \texttt{singlehop}
mode for NQ; and RAGuard retrieves top five with clean-reference \(\alpha=0.025\). All remaining parameters follow
the corresponding official implementations.

\paragraph{Compute.}
The evaluation uses up to ten NVIDIA Quadro RTX 6000 GPUs with 24\,GB of memory
each, totaling approximately 183 evaluation GPU-hours.

\end{document}